\documentstyle[12pt,twocolumn]{article}
\input epsf

\begin{document}

\begin{titlepage}

\begin{flushright}
CERN-TH 98--210\\
UM--TH  98--10\\
LAPTH  688--98\\
June 1998
\end{flushright}
\vspace{2.5cm}

\begin{center}
\large\bf
{\LARGE\bf Perturbative and nonperturbative}\\
{\LARGE\bf Higgs signals\footnote{Invited talk presented by A. Ghinculov 
           at the {\em Theory of LHC Processes} meeting, 
           9--13 February 1998, CERN, Geneva.}}\\[2cm]
\rm
{Adrian Ghinculov$^{a,b,}$\footnote{Work supported by the 
                                   US Department of Energy (DOE).}
and Thomas Binoth$^c$}\\[.5cm]

{\em $^a$Randall Laboratory of Physics, University of Michigan,}\\
      {\em Ann Arbor, Michigan 48109--1120, USA}\\[.2cm]
{\em $^b$CERN, 1211 Geneva 23, Switzerland}\\[.2cm]
{\em $^c$Laboratoire d'Annecy-Le-Vieux de Physique 
         Th\'eorique\footnote{URA 1436 associ\'ee \`a l'Universit\'e de Savoie.} LAPP,}\\
      {\em Chemin de Bellevue, B.P. 110, F-74941, 
           Annecy-le-Vieux, France}\\[3.cm]

\end{center}
\normalsize

\begin{abstract}
We discuss the current picture of the standard Higgs sector
at strong coupling and the phenomenological
implications for direct searches at the LHC. 
\end{abstract}


\end{titlepage}


\title{Perturbative and nonperturbative Higgs signals\footnote{Invited
           talk presented by A. Ghinculov at the 
           {\em Theory of LHC processes} meeting,
           9--13 February 1998, CERN, Geneva.}}

\author{Adrian Ghinculov$^{a,b,}$\thanks{Work supported by the US 
                                Department of Energy (DOE).}
       and Thomas Binoth$^c$}

\date{{\em $^a$Randall Laboratory of Physics, University of Michigan,}\\
      {\em Ann Arbor, Michigan 48109--1120, USA}\\[.2cm]
      {\em $^b$CERN, 1211 Geneva, Switzerland}\\[.2cm]
      {\em $^c$Laboratoire d'Annecy-Le-Vieux de Physique 
         Th\'eorique\thanks{URA 1436 associ\'ee \`a l'Universit\'e de Savoie.} LAPP,}\\
      {\em Chemin de Bellevue, B.P. 110, F-74941, 
           Annecy-le-Vieux, France}}

\maketitle

\begin{abstract}
We discuss the current picture of the standard Higgs sector
at strong coupling and the phenomenological
implications for direct searches at the LHC.
\end{abstract}


\vspace{.7cm}

Recently, considerable progress has been made in understanding the
nature of the standard Higgs sector when its coupling becomes strong.
Technically, computations on a lattice in the Higgs sector still have
a long way to go to attain a precision useful phenomenologically,
for instance when applied to LHC processes. Meanwhile, a new 
higher-order 
nonperturbative $1/N$ approach proved able to match the precision of
two-loop perturbative results at low coupling, while its validity extends
into the strong coupling zone as well. The availability of this nonperturbative
approach opens up the perspective to explore in a reliable way
exciting ideas such as the possibility of a Higgs boson coupled strongly
to the vector bosons and to itself, and the formation of a spectrum
of bound states at a higher scale. Such possibilities were proposed
in the past. However, they could not be worked
out from first principles because a nonperturbative solution was missing.

From the experimental point of view, should a resonance similar to
a Higgs boson be discovered at the LHC, it is crucial that its properties
be understood sufficiently well theoretically, so that a standard
Higgs can be distinguished from a nonminimal version. This can indeed
be a serious issue if everything we know is perturbation theory, as
it will become clear from the following example.

\begin{figure}[t]
\hspace{1.5cm}
    \epsfxsize = 8cm
    \epsffile{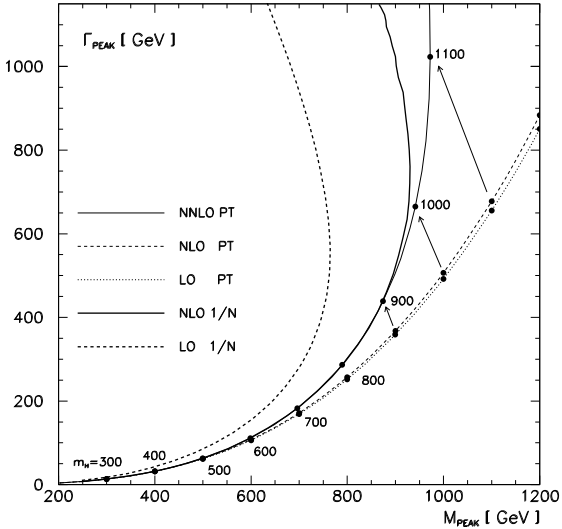}
\caption{{\em The current knowledge of the Higgs width at strong coupling. 
The mass and width parameters $M_{PEAK}$
and $\Gamma_{PEAK}$ are extracted from the position and the height 
of the Higgs resonance in fermion scattering as if the resonance was 
of Breit-Wigner type. We give the relation between $M_{PEAK}$
and $\Gamma_{PEAK}$ in perturbation theory (LO, NLO and NNLO) 
and in the nonperturbative $1/N$ expansion (LO and NLO). For
the perturbation theory curves we give the corresponding values
of the on-shell mass parameter $m_H$.}}
\end{figure}

Figure 1 summarizes the evolution of our knowledge of the width of the
Higgs boson. Here we are concerned only with the width of a heavy
Higgs boson, which decays dominantly into vector bosons. Quantum corrections
at the one-loop level were first considered in ref. \cite{marciano}. There it
was noticed that the computation of corrections of enhanced electroweak
strength can be greatly simplified by using the equivalence theorem
in Landau gauge. As one can see in this picture, the one-loop correction
turns out to be fairly small. This suggested that perturbation theory
is perfectly under control over the whole region of concern, even
well above 1 TeV. There seemed to be little point in calculating 
higher-loop corrections. 

However, at the same time another solution was known, which disagreed
numerically quite strongly with perturbation theory. This approach
attempted to calculate Green functions in the sigma model by expanding
in $1/N$, where $N$ is the number of degrees of freedom of the theory,
instead of the coupling constant. This was proposed for the 
${\cal O}(N)$--symmetric
sigma model in refs. \cite{coleman,schnitzer,dolan}. It was subsequently
applied to the standard Higgs sector in refs. \cite{casalbuoni,einhorn}.  
The resulting
leading order width does not appear to be numerically useful because
it differs substantially from perturbation theory at low coupling,
where perturbation theory is expected to be reliable. This large discrepancy
raised doubts about the consistency of the $1/N$ approach.

However, we calculated one order higher in both expansions
\cite{calc2loop,riesselmann,oneovn}, and it
turns out that the discrepancy between perturbation theory and the
nonperturbative $1/N$ expansion is reduced dramatically. As can be
seen in fig. 1, the two expansions appear to be nicely converging
towards a common solution. The next-to-leading $1/N$ solution and two-loop
perturbation theory are in a remarkable agreement up to such high
values of the Higgs mass as 800--900 GeV. 

Also it can be seen from fig. 1 that the true value of the decay width
can differ considerably from the tree and one-loop level calculations,
which so far were used widely for phenomenological studies for the LHC.
As the coupling in the Higgs sector increases, an interesting saturation
of the mass takes places, where only the total decay width grows.
Should a standard Higgs resonance be discovered at the LHC somewhere in
the zone where the saturation effect comes into place, the low-order
perturbative analysis would suggest that its coupling to the vector bosons
is too strong to be compatible with the standard model.

The two-loop perturbative analysis is based, just as the one-loop 
calculation,
on the use of the equivalence theorem in Landau gauge. The main difficulty
at the two-loop level is that it involves the evaluation of massive
two-loop Feynman graphs at finite external momentum. This is known
to be a difficult problem because the scalar integrals in the general
kinematic case are usually unknown analytic functions. One particular
case where the special functions involved were identified is the so--called
sunset self-energy topology. This was shown to be related to the Lauricella
functions \cite{lauricella}. It turns out that even in this case the diagram
is most efficiently evaluated by means of integral representations.
For this reason, we developed a general approach which is
based entirely on integral representations when the external momentum
of the graph is finite \cite{calc2loop,general2loop}. 
The case with zero external momentum
can always be treated analytically and the general solution has been
known for a long time \cite{vdbij:rho}. 
Our general solution is numerical. However,
due to the use of deterministic adaptative algorithms combined with
an optimized complex integration path defined in terms of spline functions,
the solution is fast and accurate. It was already used for the calculation
of several physical processes of phenomenological interest 
\cite{calc2loop,higgspole,gg2loop,mucollider}.
The two-loop result shown in fig. 1 was obtained with this method,
and was also reproduced in ref. \cite{riesselmann} 
with different numerical methods. 

Regarding the use of numerical versus analytical methods for this type
of calculations, we would like to make the following remark. The nontrivial
two-point functions involved in this calculation were first calculated
numerically \cite{calc2loop,riesselmann}, 
but later on an analytical solution was obtained
for them when the external momentum is on-shell \cite{jikia}. 
This was possible
because the calculation is in essence a one-scale problem if treated
in Landau gauge. This simplifies the problem considerably. For the
three-point case, because of the complexity of the diagrams, an analytical
solution was not found so far. Still, the existing numerical solution
is accurate enough for any practical purpose.

Our nonperturbative solution to this problem is based on considering
an ${\cal O}(N)$-symmetric sigma model, which recovers the 
standard model for $N = 4$. 
At an intermediate stage, a double expansion is performed, both
in the coupling constant and in $1/N$. Because of the combinatorial
structure of this theory, it is possible to calculate and to sum up
the Feynman graphs of all orders which are generated for a given order
in $1/N$. This procedure works in principle for any Green function,
but the complexity of the problem increases with the number of external
legs. 

The leading order of the $1/N$ expansion is simply the well--known geometric
series of bubble self-energy one-loop graphs and was known for a long
time \cite{coleman,schnitzer,dolan}. 
However, the next-to-leading order is much more difficult
to calculate. The first problem encountered is to identify the relevant
diagrams in all loop orders. This is most elegantly solved by a combinatorial
trick proposed by Coleman, Jackiw and Politzer \cite{coleman}. 
Their idea consists of adding a
nondynamical piece to the Lagrangian, which contains an unphysical
auxiliary field. As a result, the dynamics of the theory remains unchanged,
but the Feynman rules are modified, and there are no quartic vertices
left. This leads to a rearrangement of Feynman graphs in the higher
orders. 

Actually, this idea was used originally only at leading order, where it does
not really simplify the problem. The real power of this rearrangement
is apparent only in higher orders. For the next-to-leading order calculation
the combinatorial rearrangement of Feynman graphs is practically unavoidable.

After identifying the relevant graphs in all loop orders, a method
is needed for evaluating them. Barring a few trivial cases of limited
applicability, an analytic solution is not available. We developed
a highly efficient numerical approach based on the work of ref. \cite{3loop}
on three--loop massive graphs. In ref. \cite{oneovn} 
we applied this to two-point
functions. Meanwhile it was also extended to 
three-point functions \cite{3point}. Most
probably these methods can be extended to more complicated processes.

Because an analytical solution is not available, the ultraviolet 
$1/\epsilon$ poles
cannot be isolated from the graphs as usual and absorbed into the
$1/N$ counterterms. A few remarks about renormalization are in order
here. First, in contrast with perturbation theory, the choice 
of renormalization
scheme is of no relevance whatsoever. After summing up the complete
perturbative series of the $1/N$ coefficients no residual scheme dependence
is left. One obtains precisely the same physical result by working
in any intermediate renormalization scheme. This freedom can be best
exploited for simplifying to some extent the calculation. Second,
the wave function renormalization constants turn out to be finite,
as they should be in a nonperturbative solution. Only the coupling
constant counterterms are truly ultraviolet divergent. Since it is
complicated to extract the $1/\epsilon$ poles explicitly, 
we performed the intermediate 
renormalization in a nonstandard way, similar to the BPHZ procedure.

At the fundamental level of the theory, there is the problem of treating
the leading-order tachyons of the ${\cal O}(N)$-symmetric 
sigma model in the $1/N$
expansion. The sigma model is widely believed to be trivial, although
a rigorous proof does not exist yet. Within perturbation theory, an
indication of triviality is the existence of the Landau pole. Similarly,
in the $1/N$ expansion there is a tachyon in the Green functions. In
perturbation theory the Landau pole is generated in a region where
the beta function is not obtained reliably. Thus the Landau pole can
be considered at most an indication of triviality. In the $1/N$ expansion
the validity of the result depends only on the value of $N$, and not
on how strong the coupling is. So the previous argument does not apply.
Not much is known about the convergence properties of the $1/N$ expansion,
and it was even suggested that a nonuniform convergence may explain
the occurrence of the tachyon. Independently from what happens in
higher orders, the tachyon cannot be considered a prediction of the
theory in the usual derivation of the $1/N$ expansion. Normally the
$1/N$ solution for the Green functions is obtained by summing up its
perturbative expansion. The final result is thus determined
only up to an arbitrary function which vanishes in perturbation theory.
This freedom can be used to preserve causality. The residuum of the
tachyon pole is precisely such a function. As such, it can be subtracted
at its pole without upsetting the original information from the Feynman
diagrams. Our tachyonic regularization simply subtracts the tachyon
pole from the leading-order two-point functions. This procedure can
be repeated consistently in higher orders if necessary.

It is interesting to note that the saturation effect
is actually within the direct production reach of the Large
Hadron Collider. We only considered the standard Higgs production
by gluon fusion. The other production mechanism, the vector boson
scattering, is not yet available nonperturbatively or at two-loop order.
It is only known at one-loop \cite{dawson}.  Studies which were
performed at tree level indicate the gluon fusion to dominate up to
about 1 TeV \cite{aachenwork}. 

The gluon fusion process at hadron colliders was studied in detail
at leading order \cite{ggleading}. 
We included the correction of enhanced electroweak
strength at NNLO, as a first approximation for the nonperturbative
$1/N$ result \cite{lhcdisclim}. 
This is because the three-point function is not available
yet in the $1/N$ expansion at NLO. The use of the two-loop result is
justified up to about 1.1 TeV because the two-loop Higgs width agrees
well with the nonperturbative result. To simplify the analysis and
to avoid the need for precise detector details, such as actual energy
and angular resolutions, we confined our analysis to the nonhadronic
decay channels. We considered a 100 fb$^{-1}$ sample and we asked for a 
$5 \sigma$ effect. Then, the four charged lepton channel can reach up to an
on-shell Higgs mass of about 830 GeV \cite{lhcdisclim}. 
The two charged lepton and missing
transverse momentum channels can reach up to about 1030 GeV. As one
can see in fig. 1, this value is well within the saturation zone.
It is possible that the hadronic channels may allow one to go even
deeper in the saturation zone, due to a higher branching ratio. The
analysis is complicated by the presence of a heavy QCD background.
How well can the QCD background be separated from the signal is a
matter of detector energy and angle resolution. This is a study which
still needs to be done to assess the full potential of the LHC.

In conclusion, we now have the tools for calculating both two-loop corrections
and NLO nonperturbative $1/N$ expansions in the Higgs sector. By combining
the two expansions we were able to elucidate the strong
coupling behaviour, and to establish the presence of a mass saturation effect
at about 930 GeV. We treated the Higgs resonance in gluon fusion with
these methods, and we established that the mass saturation effect
is within the reach of the LHC even by conservatively considering only
purely leptonic channels.



\newpage


\end{document}